# Thermoelectric response of a ferroelectric insulator

Ryo Iguchi[1]*, Takashi Teranishi[2, 3], Sakyo Hirose[4], Ping Tang[5], Jun Kano[2], Gerrit E.W. Bauer[5] and Ken-ichi Uchida[1,6]

**Affiliations:**

[1]National Institute for Materials Science, Tsukuba, 305-0047, Japan.

[2]Faculty of Environmental, Life, Natural Science and Technology, Okayama University, Okayama, 700-8530, Japan

[3]Laboratory for Materials and Structures, Institute of Science Tokyo, Yokohama, 226-8503, Japan

[4] Murata Manufacturing Co., Ltd., Nagaokakyo, 617-8555, Japan.

[5]WPI-AIMR & IMR & CSIS, Tohoku University, Sendai, 980-8577, Japan

[6]Department of Advanced Materials Science, Graduate School of Frontier Sciences, The University of Tokyo, Kashiwa, 277-8561, Japan

* Corresponding author. Email: IGUCHI.Ryo@nims.go.jp

**Abstract:**

Thermoelectric effects enable the conversion between heat and electricity without moving parts. While conventionally associated with mobile charges, we report thermoelectricity caused by bound charges in the form of temperature changes measured by multi-harmonic lock-in thermography of a ferroelectric under an ac electric field. The observed temperature gradient depends on the field-induced displacement current, a Peltier effect in a dielectric material. Its coefficient exceeds 100 V around the ferroelectric-paraelectric phase transition, which is several orders of magnitude greater than reported values in conductors. Our findings uncover previously hidden functionalities of ferroelectric materials for thermal management by directional heat transport in ferroelectrics.

Thermoelectric effects are fundamental transport phenomena that enable the interconversion between heat and electricity by the Seebeck and Peltier effects, in which a temperature gradient generates voltages and charge currents are accompanied by heat currents, respectively (*1*, *2*). The belief that thermoelectric effects depend on mobile charges was overturned by the discovery of the spin Seebeck effect, in which a temperature gradient drives a spin current via the collective magnetization dynamics that does not require mobile electrons (*3*, *4*). This breakthrough launched the field of "spin caloritronics" (*5*, *6*) and the use of magnetic insulators for thermal engineering.

This success poses a compelling question: can the motion of electric polarization also drive a directed heat flow? To date, thermoelectric consequences of the polarization dynamics in dielectrics have been predicted (*7*, *8*), but never conclusively observed, indicating a lack of understanding of transport phenomena in condensed matter systems. A robust Peltier effect in insulators would allow alternative thermal management technologies with a greatly increased number of potential materials.

An electric field $E$ applied to electric insulators generates or modulates a macroscopic polarization $P$ by inducing or aligning electric dipoles, giving rise to a uniform temperature change through the electrocaloric effect (ECE), while reciprocal of the ECE, i.e., the pyroelectric effect, refers to changes in the temperature $T$ due to modulation of $P$. In the adiabatic limit, these effects can be described by equilibrium thermodynamics (*9*).

Here we address *non-equilibrium* transport phenomena, such as charge and heat currents under electric fields and temperature gradients in a planar ferroelectric capacitor. The instantaneous linear thermoelectric response in isotropic dielectrics reads

$$\begin{pmatrix} j_c \\ j_q \end{pmatrix} = \begin{pmatrix} L_{11} & L_{12} \\ L_{21} & L_{22} \end{pmatrix} \begin{pmatrix} E_{\text{eff}} \\ -\frac{\partial_x T}{T} \end{pmatrix}, \quad (1)$$

where $j_c = \partial_t P$, the time derivative of the polarization $P$, is the displacement or shift charge current density (units of $A/m^2$), $j_q$ is the heat current density, $\partial_x T$ is the gradient of the temperature $T$, and $\kappa|_{E=0} = L_{22}/T$ is a thermal conductivity. $E_{\text{eff}} = -\partial F/\partial P + E_{\text{ext}}$ is an effective field derived from the free energy $F$ of the system and an externally applied field $E_{\text{ext}} = \Delta V/d,$ where $\Delta V$ is the applied voltage, $d$ the thickness of the dielectric, and we assumed ideal screening in the metal contacts (*10*). Onsager reciprocity demands equal thermoelectric coefficients $L_{12} = L_{21}$. $j_q|_{\partial_x T=0} = j_q^{(\Pi)} = \Pi_p j_c = \Pi_p\, \partial_t P$ is the heat current driven by the displacement current or "dielectric Peltier effect" (Fig. 1) (*7*, *11*). The dielectric Peltier coefficient $\Pi_p = L_{21}/L_{11}$, while its reciprocal is the rate of change of the thermopolarization $P = \chi_T\, \partial_x T|_{E_{\text{eff}}=0}$, where $\chi_T$ denotes the thermopolarization coefficient. $\Pi_p$ has the same voltage unit V as the conventional Peltier coefficient $\Pi_{\text{c}}$ in electric conductors. The thermoelectricity of electric insulators contradicts intuition, and previous attempts of experimental verification have not been conclusive (*12*, *13*).

Here, we report the observation of a transient Peltier effect in a ferroelectric insulator by multi-harmonic lock-in thermography (LIT) (Fig. 2) (*14–16*) that enables clear separation of Peltier and ECE signals at different harmonic frequencies under an alternating electric field. We analyze the results by the Onsager phenomenology of a dissipative dielectric. The Peltier coefficient $|\Pi_p| > 100$ V is many orders of magnitude larger than the metallic Peltier ones in electronic and ionic thermoelectric materials (*17*, *18*).

We chose the well-understood material $BaTiO_3$ (*19–21*) as a parent compound for our studies. Its phase transition can be tuned into the temperature range accessible to our equipment (300–390 K) by alloying it with Sr ($Ba_{0.8}Sr_{0.2}TiO_3$) (*22–24*) and a small concentration of a compensating donor (see materials and methods). We fabricated a planar capacitor by depositing Au electrodes on the edge faces of rectangular bars. We applied a slowly varying electric field $E_{\mathrm{ext}}(t) = E_0 \sin(2\pi f t)$ with $f \in [0, 60]$ Hz with amplitude $E_0$ that is large enough to fully revert the polarization $P$ during each cycle (Fig. 2B). The LIT imaged the resulting temperature distribution at a bare surface as illustrated by Fig. 2C. The frequency harmonics of the infrared signal detect different thermal effects (Figs. 2D–F): The ECE is sensitive to the magnitude but not the direction of $P$ and therefore oscillates with frequency $2f$. In contrast, the heat current and the resultant temperature gradient generated by the dielectric Peltier effect is proportional to $E_{\mathrm{eff}}(t)$ and contributes to the first harmonic ($1f$). While the ECE only causes uniform temperature changes (*16*, *25*), the sample edges are heat sinks and sources due to the finite divergence of the Peltier heat current (Fig. 2E), creating non-uniform temperature profiles governed by heat diffusion into the sample.

Figures 2E and F expose the thermal LIT profiles at the first ($1f$) and second ($2f$) harmonic under $E_0$ = 1 kV/mm, $f$ = 10 Hz, and $T$ = 300 K in terms of the amplitude $A_{nf}$, the absolute magnitude of the induced temperature change, and the phase delay $\varphi_{nf}$ that characterize the time lag of the temperature modulation (*14*). The first harmonic ($1f$) signal is localized at the left and right edges close to the electrodes. Their opposite sign indicates heating of one edge and cooling of the other—the hallmark of the Peltier effect (*26*). On the other hand, the second harmonic ($2f$) response caused by the electrocaloric effect (ECE) is uniform across the entire surface as expected (*16*).

We measure the time-dependent polarization $P(t)$ by a Sawyer-Tower circuit (shown in Fig. 2G) (*27*) simultaneously with the thermographic images. The observed $P$-$E$ hysteresis in Fig. 2G proves that the material is ferroelectric at 300 K with a small coercivity of 0.11 kV/mm. Its point symmetry with respect to the origin indicates that the $1f$ signals are not contaminated by residual poling effects. $P(t)$ oscillates mainly at frequency $1f$, with higher harmonic components accounting for less than 0.1 %. Moreover, the third and higher harmonic components in the LIT signals are very weak.

The driving frequency dependence of the thermoelectric response provides additional clues about the physical mechanism. According to Fig. 3A, the Peltier effect increases and becomes more localized at the edges with increasing $f$. We analyze the response in terms of the observed edge-to-edge temperature difference $\Delta T_{\mathrm{edge}}^{(1f)} \exp\left(i\varphi_{\mathrm{edge}}^{(1f)}\right) = A_{\mathrm{R}} \exp(i\varphi_{\mathrm{R}}) - A_{\mathrm{L}} \exp(i\varphi_{\mathrm{L}})$, where the subscripts L and R denote the $1f$ signals averaged over 10 pixels (170 μm) at the left and right edges, respectively (see Fig. 3A). $\Delta T_{\mathrm{edge}}^{(1f)} \rightarrow 0$ at low frequencies ($f$ < 1 Hz) in Fig. 3B signals the absence of leakage currents due to doping or thermally excited carriers (*28–31*). Since the $P$-$E$ hysteresis loop in Fig. 3C does not depend on frequency (see the inset to Fig. 3C), we attribute the linear increase of the temperature with $f$ at low frequencies to the Peltier effect $\Delta T_{\mathrm{edge}}^{(1f)} \sim f$, very different from the dc Peltier effect in metals (see Refs. (*32*, *33*)). The increase of the localization at the edges with $f$ as well as saturation for $f$ > 30 Hz can be understood by thermal diffusion (see below). The independence of the phase-delay $\varphi_{\mathrm{edge}}^{(1f)}$ on frequency points to a robust appearance of the temperature difference.

We derive the magnitude ($q_\Pi$) and phase ($\varphi_\Pi$) of the induced heat current density $j_q^{(\Pi)} = q_\Pi \exp(-i\varphi_\Pi)$ by a two-parameters fit to the first harmonic of the observed temperature changes via the one-dimensional heat diffusion equation. The heat sources $-\partial j_q^{(\Pi)}$ are delta functions at the interfaces to the contacts. The temperature profiles are more localized near the edges at higher frequencies because the energy has less time to escape the sources. The fitted profile and the experimental data of both the in- ($\Delta T_{\text{edge}}^{(1f)} \cos \varphi_{\text{edge}}^{(1f)}$) and out-of-phase ($\Delta T_{\text{edge}}^{(1f)} \sin \varphi_{\text{edge}}^{(1f)}$) components in Figs. 3D and 3E agree for all frequencies, validating our interpretation and the accuracy of the extracted $q_\Pi$ and $\varphi_\Pi$ values in Fig. 3G.

We now turn to the complex Peltier coefficient in frequency space $\Pi_p = (q_\Pi/j_c^{(1f)}) \exp(-i[\varphi_\Pi - \varphi_c^{(1f)}])$. The real part measures the thermoelectric conversion while the delay between the induced heat current and the displacement current governs the imaginary component. Using the $1f$ components of the displacement current $j_c^{(1f)}$ and $\varphi_c^{(1f)}$ in Fig. 3H, we find $\Pi_p \approx -28.3 + 9.4\text{i V}$ for $f \geq 10$ Hz. $\text{Re}(\Pi_p) < 0$ indicates that the heat flows in the opposite direction of the charge current $j_c$. $|\Pi_p| \sim O(10\ V)$ is three orders of magnitude larger than the typical Peltier coefficients $\Pi_c \sim O(10\ \text{m}V)$ for metals and semiconductors (*17*)! The imaginary part indicates that the heat current is significantly phase-delayed relative to the displacement current by ~16 degrees, indicating the role of a slower degree of freedom such as the motion of domain walls in the conversion process.

Next, we confirm the intimate relation between the thermoelectric response and dielectric fluctuations. We focus on the temperature range across the diffuse ferroelectric-to-paraelectric phase transition, which is characterized by the temperature that maximizes the dielectric response at $T_m \approx 340$ K. The raw LIT images reveal strongly enhanced $1f$ (Peltier) and $2f$ (ECE) signals near $T_m$ (Figs. 4A and 4B) but with characteristic differences. Figures 4C and 4D show the $T$ dependence of $\Delta T_{\text{edge}}^{(1f)}$ from the $1f$ response, and the bulk temperature change, $\Delta T_{\text{avg}}^{(2f)}$ averaged over the sample surface, respectively. The pronounced peaks near $T_m$ confirm the pivotal role of critically enhanced dielectric fluctuations near the phase transition on the thermoelectric properties. The $\Delta T_{\text{edge}}^{(1f)}$ peak is significantly broader than the sharp $\Delta T_{\text{avg}}^{(2f)}$ peak. Moreover, the Peltier response ($\Delta T_{\text{edge}}^{(1f)}$) persists over a larger temperature range than the ECE, presumably because of the different microscopic mechanisms.

A detailed data analysis allows us to distill the key observables $q_\Pi$, $j_c^{(1f)}$, and $\Pi_p$ corrected for the temperature dependence of the material's thermal parameters. Above $T_m$, the hysteresis of the *P*-*E* curves vanishes with reduced susceptibility (Fig. 4E), consistent with the broad cusp of $j_c^{(1f)} = \partial_t P$ at $T_m$ (Fig. 4F). The enhanced heat current around $T_m$ (Fig. 4G) indicates the critical enhancement of the transport dielectric Peltier coefficient $\Pi_p$ observed in Fig. 4H.

We further confirmed the intrinsic, bulk origin of the observed thermoelectric response, we systematically investigated its dependence on electrode and probe materials and found no significant differences. Rotated-sample measurements with spatially resolved detection further demonstrated that the thermoelectric heat current is generated throughout the bulk of the material (Fig. 5). Additionally, a history-dependent hysteresis in the dielectric Peltier response was observed below ($T_m$), with the response increasing and stabilizing after repeated measurements.

The observed thermoelectric response differs from that observed in conducting ferroelectrics (*29–31*) and in the paraelectric phase of insulating ferroelectrics (*12*, *13*). Our sample is highly insulating, with an electrical conductivity less than $10^{-8}$ S/m, while the vanishing temperature gradient in the dc limit demonstrates the irrelevance to the conventional thermoelectric response due to mobile charges. Previous reports on a Peltier response of insulating ferroelectrics differ in terms of magnitude, temperature dependence, and sign. In $Pb(Mg_{1/3}Nb_{2/3})O_3$ (PMN) single crystals with $T_m \sim 263$ K, $\Pi_p \sim 0.9$ V at $E \sim 1$ kV/mm was found in the paraelectric phase $T = 270$–$320$ K (*12*, *13*). Moreover, in those experiments the electrode in contact with the positive end of the polarization grew hot during charging ($\partial_t P > 0$) while we observe cooling ($\Pi_p \partial_t P < 0$). On the other hand, a reported thermopolarization (*34*–*39*) agrees in terms of the sign ($\chi_T = -\chi_p \Pi_p / T > 0$ when disregarding the small imaginary part), where $\chi_p = P/E_{\text{ext}} > 0$ is the susceptibility. However, the reported values are very small ($\chi_T \sim 10^{-12}$C/m/K) compared with $\chi_T \sim 10^{-8}$ C/m/K reported here. Those studies focused on the paraelectric phase (without external electric fields) and therefore missed a possible critical enhancement near $T_m$.

We derive the dynamical linear response of a ferroelectric to the field $E_{\text{ext}}(t) = E_0 \sin(2\pi f t)$. Since the experiments fully revert the polarization with each cycle, this analysis is valid only close to the critical temperature. For small polarization amplitudes $|P(t) - P_0|$ from the equilibrium value $|P_0|$, the overdamped Landau-Tani-Khalatnikov equation can be linearized to the equation of motion

$$\partial_t P + \frac{P(t) - P_0}{\tau} = \frac{\chi_p}{\tau} E_{\text{ext}}(t), \tag{2}$$

where $\tau$ is the intrinsic polarization relaxation time. The displacement current then reads

$$j_c(t) = \frac{2\pi f \chi_p E_0}{\sqrt{1 + 4\pi^2 f^2 \tau^2}} \cos(2\pi f t - \phi). \tag{3}$$

where $\phi = \tan^{-1}(2\pi f \tau)$ is a relaxation phase delay. $j_c(t) \sim f$ without observable saturation in the experiments (Fig. 3C) implies that $f\tau \ll 1$. This result is consistent with relaxation times in ferroelectrics of the order of nano- to picoseconds (*23*).

Without thermal diffusion and adopting $f\tau \ll 1$, the calculated temperature gradient is constant over the sample

$$\partial_x T^{(1f)} = \frac{\Pi_p - \Pi_c}{\kappa} 2\pi f \chi_p E_0 \cos(2\pi f t) \tag{4}$$

Equation (4) only governs the order of magnitude of the temperature changes. In the supplementary text, we explain that they are localized to the edges on the scale of the frequency-dependent heat diffusion length. The susceptibility $\chi_p = P/E_{\text{ext}}$ in Eq. (3) explains the critically enhanced ECE effect $\sim \partial_T \chi_p$ in Fig. 4D at $T_m$, but only partly the peak in Fig. 4C, since $\Pi_p$ also strongly varies (Fig. 4H). Finally, $j_c^{(1f)} \propto f \chi_p E_0$ in Fig. 4F also increases around $T_m$, in contrast to the thermal conductivity $\kappa$ that is dominated by acoustic phonons (*40*).

The microscopic mechanism for the ac dielectric Peltier effect is presently not established. The displacement current in a polar material can be associated with a heat current when the dynamics are coupled to a reservoir with energy-dependent density of state, but we are not aware of models that explain the order of magnitude of the observed $\Pi_p$. The critical enhancement when approaching the phase transition from the paraelectric phase shows that broken inversion symmetry is not required but it rather reflects importance of the large fluctuations unique to the soft phonon or ferron modes in ferroelectrics (*8*, *41*).

We should exercise caution in interpreting the results since $\Pi_p$ depends on the driving field amplitude $E_0$ as shown in Fig. 6, indicating that we operate in a non-linear regime while the above analysis assumes linear response. We may speculate that the switching dynamics of the ferroelectric texture should play a role. In the ferroelectric phase, polarization is reversed by the nucleation and subsequent propagation of domain walls to grow domains with polarization aligned to the electric field. The domain walls in a planar ferroelectric are predicted to move against an applied temperature gradient regardless of the polarization texture of the walls (*42*). The reciprocal effect demands that moving domain walls generate heat currents. The observed sign of $\Pi_p$ and the phase delay of the thermoelectric response appears to be consistent with this picture, especially considering that the domain walls require 1 ms to 1 μs to propagate through a 1 mm thick sample (*43*, *44*).

Our findings have three major implications. First, the Peltier response in ferroelectric insulators offers an additional degree of freedom for the design of thermoelectric controllers, which conventionally rely on mobile charges and spins. There are many solutions to rectify an ac response for steady state heating and cooling applications, developed for conventional electrocaloric and magnetocaloric technologies (*45–49*). Thus, along with the large Peltier coefficient for ferroelectric materials, the dielectric Peltier effect may facilitate alternative strategies for improved thermal management devices. Second, our results suggest that the dynamics involved in polarization reversal plays an important role in the heat flow in dielectric insulators, much like dynamics of elementary spin and polarization excitation affects the thermal properties of magnets (*50–52*) and ferroelectrics (*53*, *54*). This may ultimately lead to a better understanding of energy transport in ferroelectric materials. Third, we suggest revisiting prior electrocaloric studies in capacitor samples: the Peltier-like response is intertwined with the ECE's bulk temperature change, for example when only one surface of a planar capacitors is measured in the time-domain (see rotated-sample measurements in Fig. 5). Disentangling the reversible and irreversible contributions should be crucial for assessing ECE performance and for designing more efficient caloric devices that may synergistically exploit both effects.

In conclusion, our study reveals a previously overlooked thermoelectric response intrinsic to ferroelectric polarization dynamics. The dielectric Peltier effect is largest at the ferroelectric transition temperature and exhibits nonlinear dependencies on external parameters. We cannot offer a microscopic understanding but suspect that the observed heat currents are driven by domain wall motion. Our findings establish the dynamic polarization of ferroelectric materials as a key carrier of energy transport in insulators.

## Materials and Methods

### Material

The Sr-substituted $BaTiO_3$ was prepared through solid-state reaction and conventional tape casting methods. A mixture of high-purity $BaCO_3$, $SrCO_3$, $TiO_2$, and $MnCO_3$ powders with nominal composition of $Ba_{0.8}Sr_{0.2}Ti_{0.998}O_3$ was milled in distilled water using partially stabilized zirconia balls for 16 hours. To improve insulation resistance, we substituted 0.2% of Ti by Mn. The resulting slurry was dried and calcined at 1150°C for 2 hours, ball-milled in an organic solvent with binder for 16 hours and fabricated into green sheets by the doctor blade method. Green samples were obtained by stacking, pressing, and cutting the green sheets. The binder was removed by heating at 500°C for 24 hours, and the samples were finally sintered at 1350°C for 4 hours in air atmosphere.

The fabricated samples were characterized using a scanning electron microscope (SEM, ERA-8900, ELIONIX Inc.), X-ray difraction (XRD, SmartLab 3, Rigaku) and impedance analyzer (Novocontrol Tech-nologies). Additional resistance measurements using an electrometer (Keithley Instruments, Model 6514) showed that the resistance of the sample used in the lock-in thermography measurements exceeds 200 GΩ for the measurement temperature range of 300 to 400 K, corresponding to a resistivity greater than > $6 \times 10^8$ Ωm.

Lock-in thermography

For the lock-in thermographic measurements, the sample was fabricated as a rectangular bar ($1.0 \times 1.5 \times 2.0$ mm$^3$). On the edge faces ($1.5 \times 2.0$ mm$^2$), a gold electrode with a thickness of about 100 nm were sputter coated in air. The sample was placed on top of double-sided polyimide tape to ensure only weak thermal contact with the variable temperature stage, thus allowing the sample to be treated as an isolated system. An electric field, $E_{\mathrm{ext}}(t)$, was applied across these electrodes via thin spring probes to drive cyclic polarization reversals. The LIT measurements imaged the resulting temperature distribution on the bare surface between the electrodes as shown in Fig. 2C. According to the thermographic image, the actual dimensions of the observed surface are estimated to be $0.95 \times 1.50$ mm$^2$.

The lock-in thermographic measurements are performed using an infrared camera (InfraTec GmbH, VarioCAM HD 800 with microscopic 1× lens) with a 7.5–14 μm detection band and a spatial resolution of 17 μm per pixel. The camera continuously sent the temperature distribution of $1024 \times 96$ pixels at a rate of 240 Hz to a processing unit, where the thermal images are decomposed into multi-harmonic frequency components (16). Here, the maximum lock-in frequency is limited to the one fourth of the camera frequency to reliably extract in-phase and out-of-phase components. The LIT images shown in the main text are obtained by cropping and correcting the rotation of the sample. The rotation angle was determined based on edge detection, and the images were realigned by applying bilinear interpolation separately to the in-phase and out-of-phase components. The measured camera values on the bare surface of the sample were calibrated using the actual temperature dependence of the camera values, as detailed in Ref. (16), as well as the time constant calibration.

Polarization measurements

The temporal evolution of polarization dynamics was measured using a reference capacitor of 0.1 μF and a reference resistor of 10 MΩ (27). The voltage across the reference circuit was monitored with a data acquisition unit, together with a trigger signal synchronized to the electric field. The values of the polarization $P$ and the displacement current $j_{\mathrm{c}}$ values were obtained by calculating the impedance of the reference circuit at each frequency and considering the sample dimensions.

Extraction of thermoelectric parameters

To extract the amplitude $q_\Pi$ and phase $\varphi_\Pi$ of the Peltier-induced heat current $j_q^{(\Pi)}$, we fit the profile of the first harmonic LIT image along the $T$-axis in the direction between the electrodes. The profile is obtained by averaging the image over the sample along the direction perpendicular to the $T$-axis in the image. The fitting function is given by

$$A_{1f}(x)e^{-i\varphi_{1f}(x)} = \frac{1-i}{2}\frac{\lambda}{\kappa}\operatorname{sech}\left(\frac{1+i}{\lambda}\frac{d}{2}\right)\sinh\left(\frac{1+i}{\lambda}x\right)q_\Pi e^{-i\varphi_\Pi} \quad (5)$$

where $A_{1f}$ and $\varphi_{1f}$ represent the amplitude and phase delay of the LIT signal at the first harmonic, respectively, $\kappa$ is the thermal conductivity, and $\lambda$ is the distance between the electrodes. $\lambda = \sqrt{\alpha/\pi f}$ is the thermal diffusion length, where $\alpha$ is the thermal diffusivity. In the fitting, $\kappa$ and $\alpha$ are fixed to the values obtained by spline interpolation of the experimental data, while $q_\Pi$ and $\varphi_\Pi$ are treated as free parameters.

**Acknowledgments**

R.I. and K.U. thank J.P. Heremans for fruitful discussions and M. Isomura and D. Fukuda for technical assistance. This work was supported by JSPS KAKENHI Grant-in-Aid for Scientific Research (B) (20H02609) (R.I., T.T., J.K., K.U.), JSPS KAKENHI Grant-in-Aid for Scientific Research (S) (22H04965) (R.I., G.B., K.U.), JSPS KAKENHI Grant-in-Aid for Scientific Research (B) (24K01334) (R.I., T.T.), JSPS KAKENHI Grant-in-Aid for Scientific Research (B) (26K00625) (P.T.), JSPS KAKENHI Grant-in-Aid for Scientific Research (B) (24K01162) (T.T.), JST ERATO “Magnetic Thermal Management Materials” (JPMJER2201) (K.U.),JST CREST (JPMJCR25S4) (T.T.), and NIMS Joint Research Hub Program.

### Author contributions

Conceptualization: R.I., G.B., K.U.; Methodology: R.I., P.T., G.B., K.U.; Resources: R.I., T.T., S.H., K.U.; Investigation: R.I., T.T., S.H.; Visualization: R.I.; Funding acquisition: R.I., T.T., P.T., J.K.,G.B., K.U.; Project administration: R.I., K.U.; Supervision: R.I., K.U.; Writing – original draft: R.I., G.B.; Writing – review and editing: T.T., S.H., J.K., P.T., K.U.

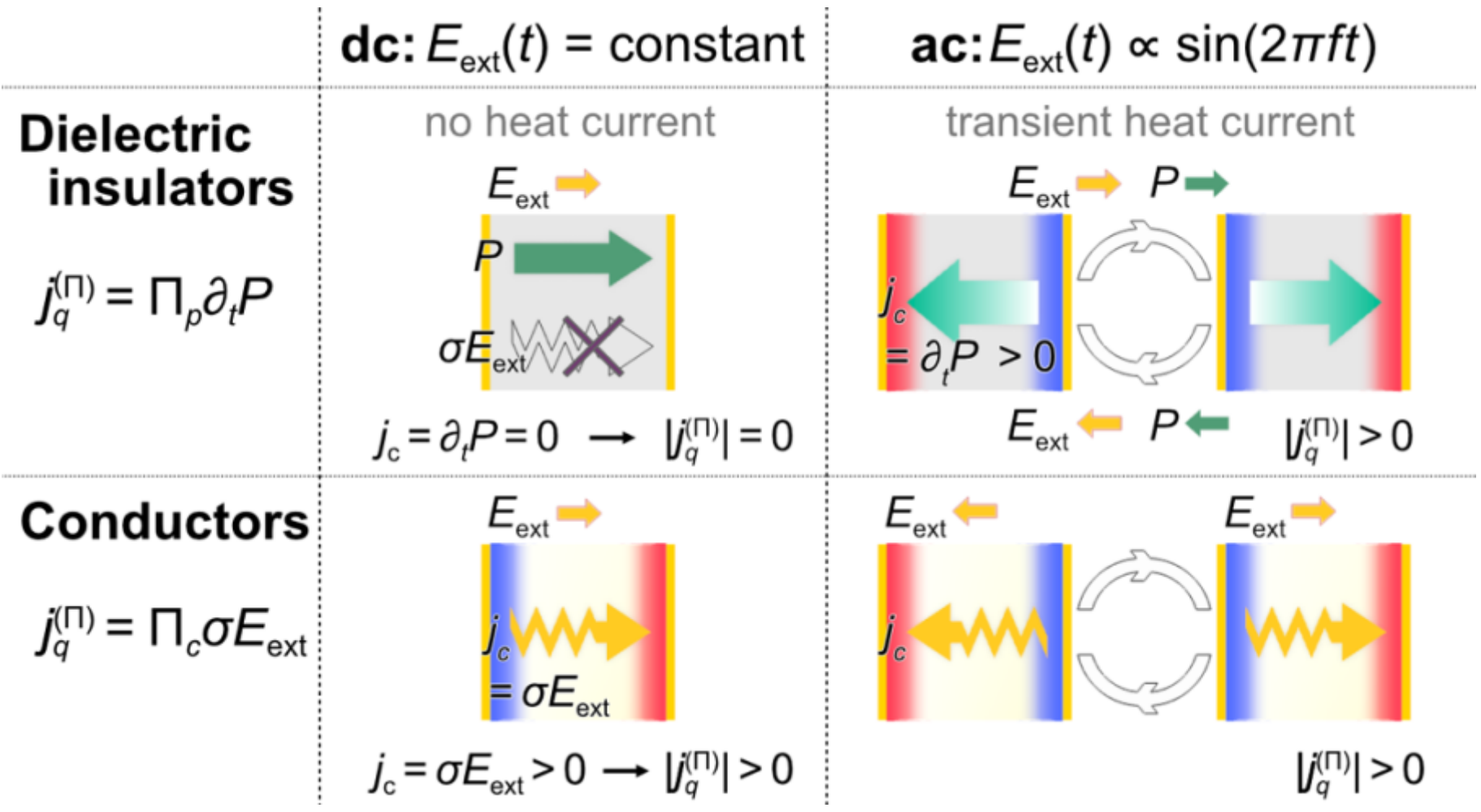

**Fig. 1. Peltier response to ac and dc electric fields in dielectric insulators and conductors, respectively.** In conductors, the application of $E_{ext}$ induces both charge ($j_c = \sigma E_{ext}$) and Peltier heat ($j_q^{(\Pi)}$) currents for ac and dc electric fields, where $\sigma$ denotes the conductivity. In dielectric insulators, a static external electric field ($E_{ext}$) causes a static polarization ($P$) but vanishing displacement ($j_c = \partial_t P$) and Peltier heat ($j_q^{(\Pi)}$) currents. However, under an ac electric field, the polarization as well as displacement ($j_c$) and heat ($j_q^{(\Pi)}$) currents oscillate with finite amplitude.

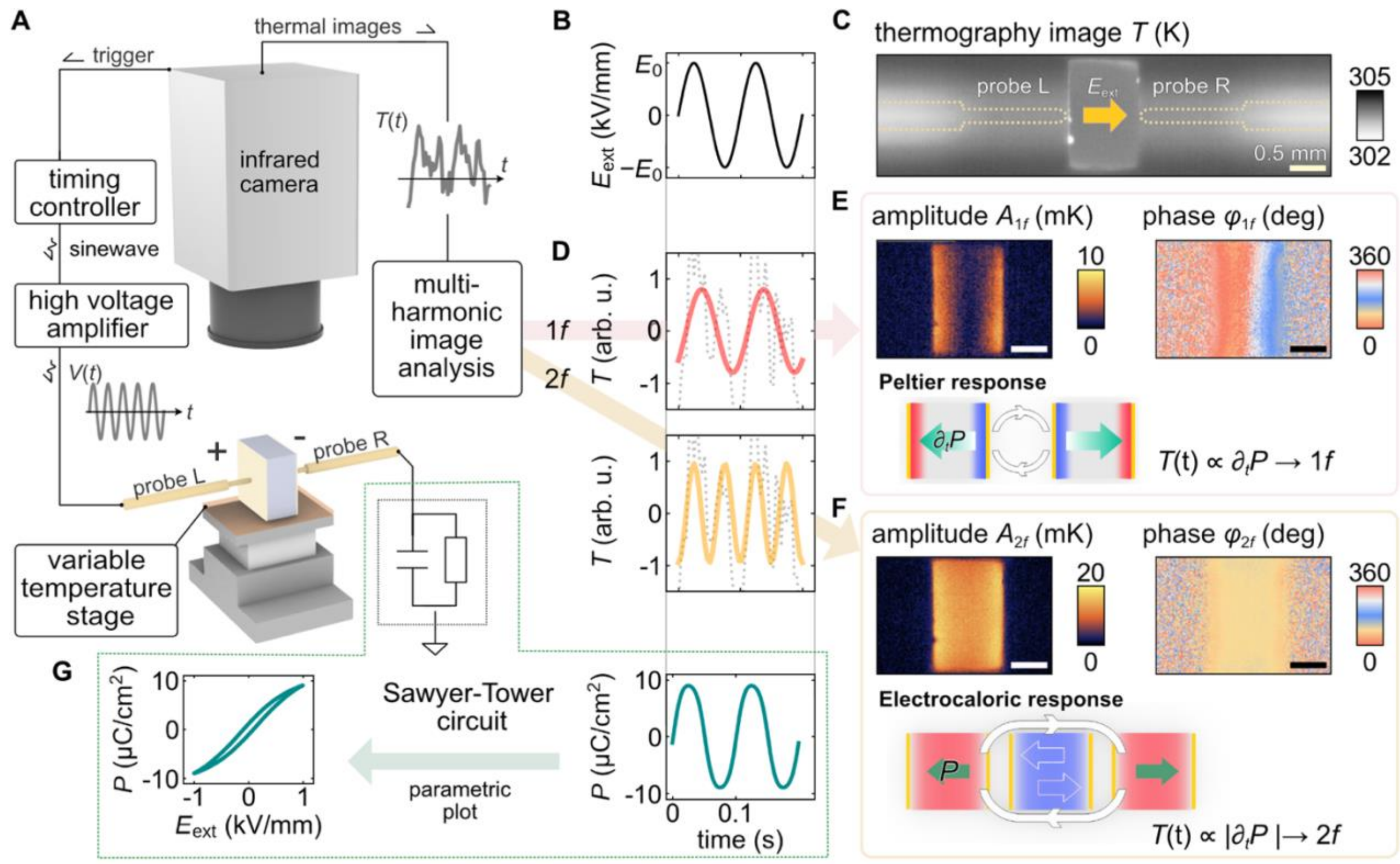

**Fig. 2. Schematic of experimental setup and thermal images.**

(**A**) Experimental setup for observing the temperature changes in a ferroelectric capacitor induced under an oscillating electric field $E_{\mathrm{ext}}(t)$ by multi-harmonic lock-in thermography (LIT). (**B**) Temporal evolution of $E_{\mathrm{ext}}(t)$. (**C**) Thermographic snapshot showing the sample surface and probes. (**D**) First (1*f*) and second (2*f*) harmonics of the time-dependent temperature $T(t)$. (**E, F**) LIT images at 1*f* and 2*f*. The white and black bars in the images indicate a length of 0.5 mm. (**G**) Left panel: $P$-$E_{\mathrm{ext}}$ hysteresis measured and synthesized by a Sawyer-Tower circuit integrated into the LIT setup. Right panel: raw time dependent $P(t)$.

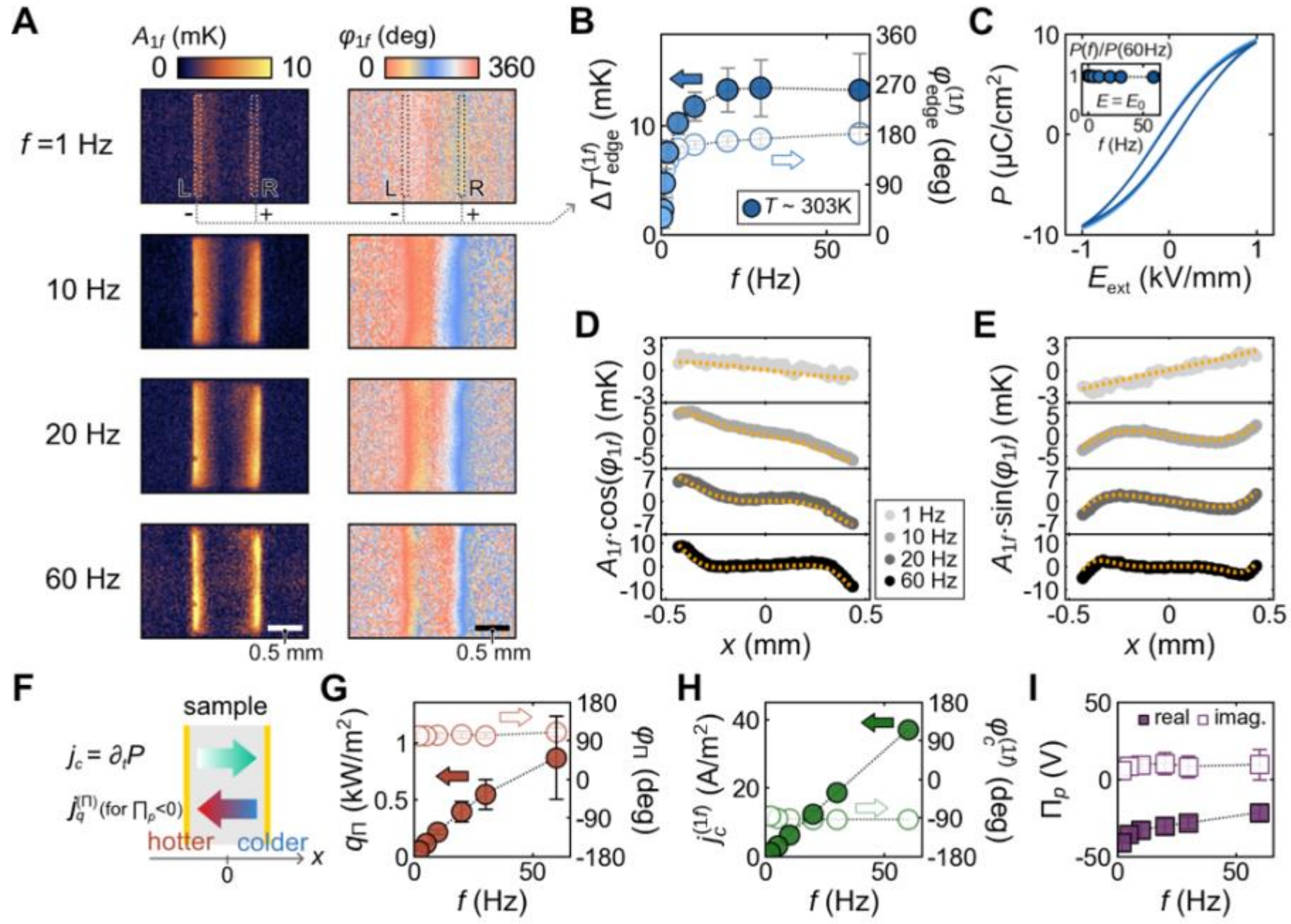

**Fig. 3. Frequency dependence of the Peltier response at room temperature.**

(**A**) First harmonic LIT images at different excitation frequencies (*f*). The dotted line marks the region used for calculating the edge-to-edge temperature difference, $\Delta T_{\mathrm{edge}} e^{-i\varphi_{\mathrm{edge}}}$. (**B**) *f* dependence of the amplitude ($\Delta T_{\mathrm{edge}}$, left axis) and phase ($\varphi_{\mathrm{edge}}$, right axis) of the temperature difference, plotted as filled and open circles, respectively. (**C**) The *P*-*E* loops for different *f* overlap. The color of each curve indicates the measured frequency, as defined in (B). The inset shows $P(E_0)$ as a function of *f*. (**D, E**) In-phase (D) and out-of-phase (E) components of the temperature change along the *x*-axis. The profiles are averaged over the sample height in the vertical direction of the image, and the outermost pixels are excluded to avoid the influence of edge defects. The yellow dotted lines are the best-fit curves, assuming Peltier-type heat sinks and sources at sample edges. (**F**) Heat and displacement current directions. (**G**) Extracted values of the Peltier-induced heat current amplitude ($q_{\Pi}$) and phase ($\varphi_{\Pi}$). (**H**) Extracted values of the displacement current density at the first harmonic ($j_c^{(1f)}$) and its phase ($\varphi_c^{(1f)}$). (**I**) Fitted complex Peltier coefficient $\Pi_p(f)$.

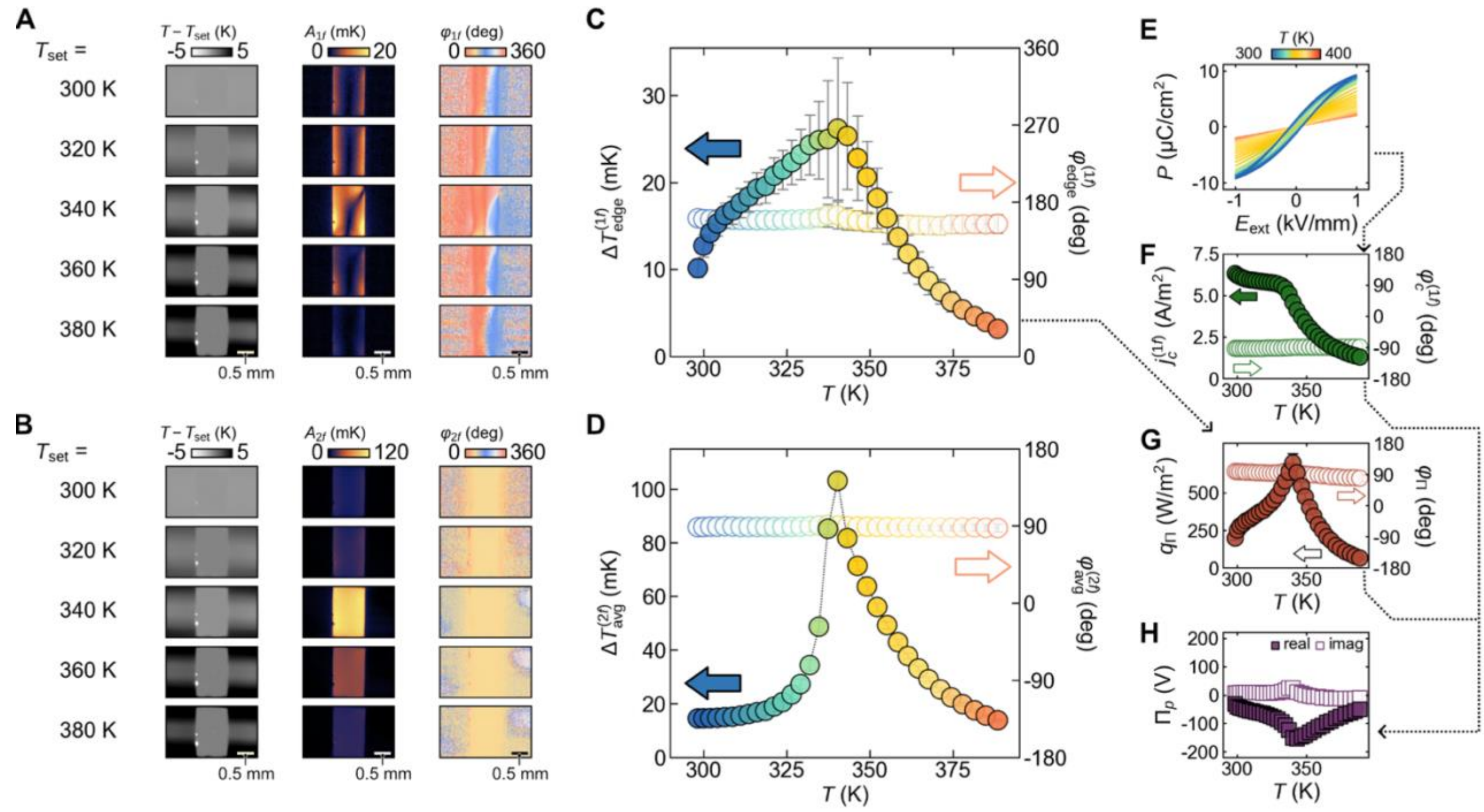

**Fig. 4. Temperature dependent of Peltier (1*f*) and electrocaloric (2*f*) response**.

(**A, B**) 1*f* (A) and 2*f* (B) components of the LIT images at $T \sim 300$, 320, 340, 360, and 380 K and $f = 10$ Hz. The 1*f* images display the Peltier induced temperature gradients, while the 2*f* images show the uniform temperature change induced by the electrocaloric effect. (**C**) *T* dependence of the amplitude ($\Delta T_{\mathrm{edge}}$, left axis) and the phase ($\varphi_{\mathrm{edge}}$, right axis) of the temperature difference, plotted as filled and open circles, respectively. (**D**) *T* dependence of the amplitude ($\Delta T_{\mathrm{avg}}^{(2f)}$, left axis) and phase ($\varphi_{\mathrm{avg}}^{(2f)}$, right axis) of the temperature changes averaged over the sample surface at 2*f*, plotted as filled and open circles, respectively. (**E**) $P$-$E_{\mathrm{ext}}$ curves measured at various temperatures. The colors are identical to those of the filled circles in panels (C) and (D), that code the temperatures. (**F**) *T* dependence of the displacement current density $j_c^{(1f)}$ and phase $\varphi_c^{(1f)}$ at the first harmonic frequency. (**G**) *T* dependence of the extracted amplitude of the Peltier induced heat current $q_\Pi$ (filled circle, left axis) and its phase $\varphi_\Pi$ (open circle, right axis) (**H**) *T* dependence of the Peltier coefficient $\Pi_p$.

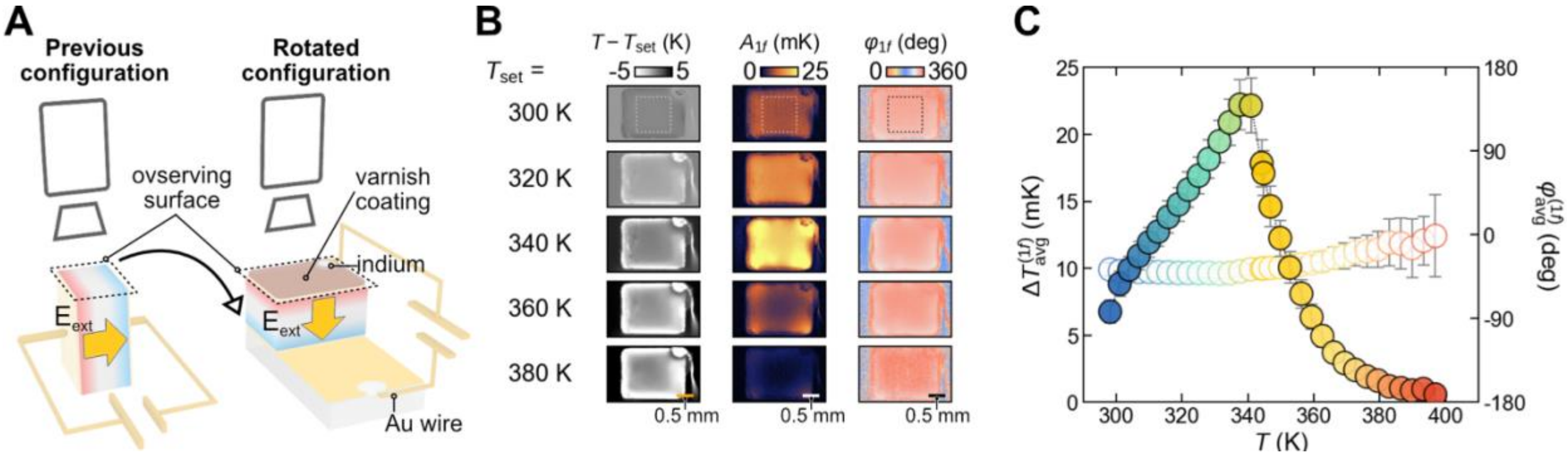

**Fig. 5. LIT measurements of a coated rotated sample.**

(**A**) Schematic to measure the Peltier-induced temperature changes over the contact. The quartz glass plate mount acts as a thermal bath for the bottom electrode. (**B**) Spatially resolved 1*f* LIT images measured at $T\sim 300$, 320, 340, 360, and 380 K with $f = 10$ Hz. (**C**) Temperature change $\Delta T_{\text{avg}}$ and $\varphi_{\text{avg}}$ averaged over the entire electrode surface at 1*f*. Its values are reduced due to the heat leakage to the mount and suppressed thermal diffusion through the top coating. However, the overall as a function of temperature is the same in both configurations.

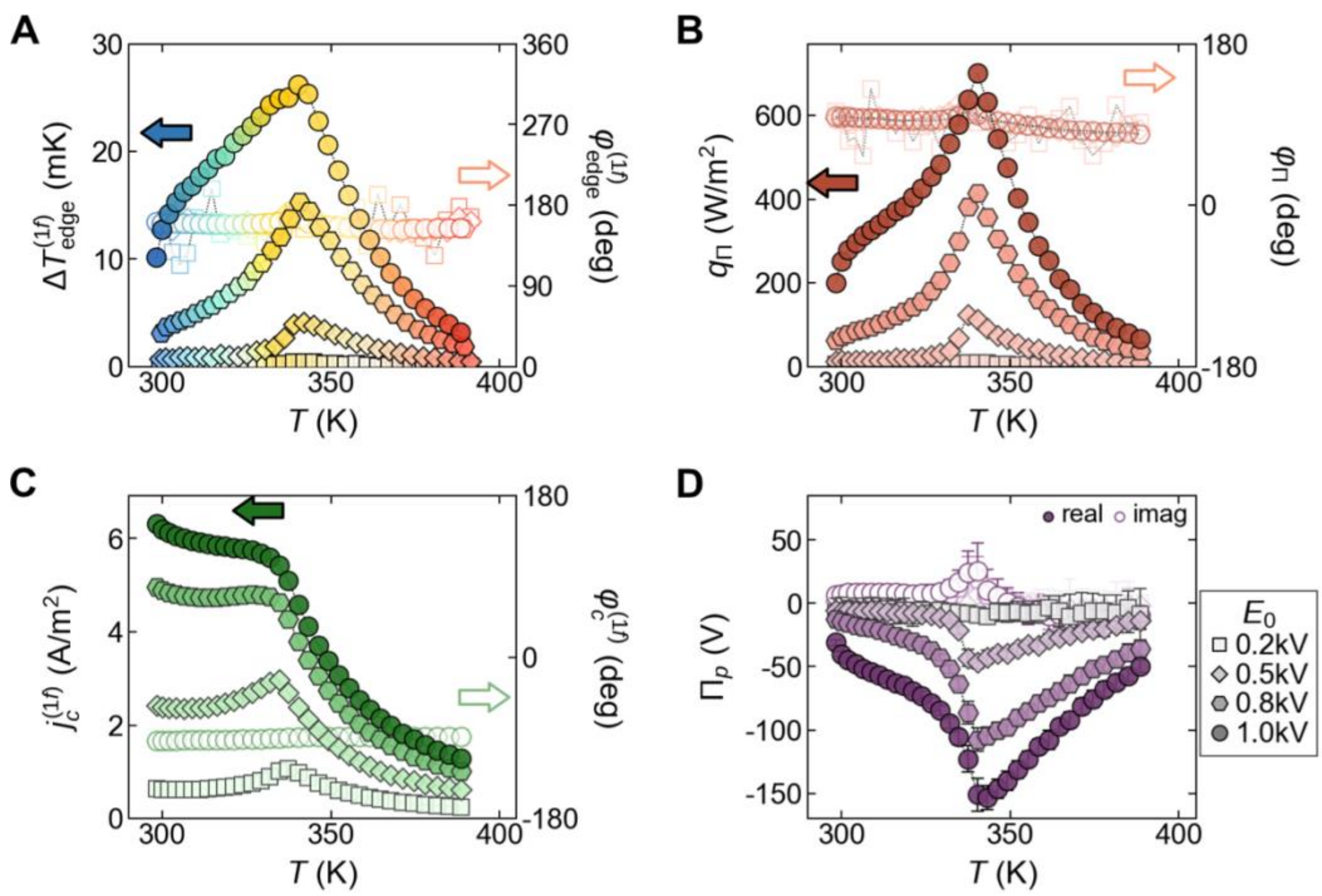

**Fig. 6. Effect of drive amplitude on the Peltier response.**

(**A**) *T* dependence of the amplitude $\Delta T_{\text{edge}}$ (filled circle, left axis) and the phase $\varphi_{\text{edge}}$ (open circle, right axis) for different drive amplitudes ($E_0$) (coded by the markers defined in the legend of panel D). (**B**) *T* dependence of $q_\Pi$ and $\varphi_\Pi$ for different $E_0$ drives. (**C**) *T* dependence of $j_c^{(1f)}$ and $\varphi_c^{(1f)}$ for different $E_0$ drives. (**D**) *T* dependence of $\Pi_p$ for different $E_0$ drives.